\documentclass[aps,pra,showpacs]{revtex4}
\usepackage{amsmath}
\usepackage{graphicx}
\usepackage{dcolumn}
\usepackage{longtable}

\newcommand{\be}{\begin{equation}}
\newcommand{\ee}{\end{equation}}
\newcommand{\bearray}{\begin{eqnarray}}
\newcommand{\eearray}{\end{eqnarray}}
\newcommand{\bse}{\begin{subequations}}
\newcommand{\ese}{\end{subequations}}

\begin{document}

\title{Higher order corrections to the hydrogen spectrum from the Standard-Model Extension}

\author{Theodore J. Yoder}
\author{Gregory S. Adkins}
\email[]{gadkins@fandm.edu}
\affiliation{Franklin \& Marshall College, Lancaster, Pennsylvania 17604}
\date{\today}

\begin{abstract}
We have studied the effects of the Standard-Model Extension (SME) on hydrogen as a realization of new physics effects that incorporate Lorentz and CPT violation.  Specifically, we calculated the SME-induced energy level shifts at order $\alpha^2$ times the SME parameters.  We obtained contributions at this order both from the non-relativistic effective Hamiltonian for motion of a spin-1/2 particle in the presence of SME interactions and also from SME corrections to the propagator for exchange photons.  We applied our result to the $2S-1S$ transition in hydrogen, which has been measured with extremely high precision.  The results obtained in this work give the leading SME corrections for this transition.
\end{abstract}


\maketitle

%%%%%%%%%%%%%%%%%%%%%%%%%%%%%%%%%%%%%%%%%%%%%%%%%%%%%
% version 7

\section{Introduction}
\label{introduction}

For the past century atomic hydrogen has played a unique role at the intersection between quantum theory and experiment.  One need only think of the Bohr model, Schr\"odinger's wave mechanics, hydrogen fine structure, the Lamb shift, the hydrogen maser, hydrogen condensates, the importance of hydrogen in astrophysics and cosmology, the multitude of ultra precise spectroscopic measurements, and hydrogen's use in the least squares adjustment of the fundamental constants \cite{Mohr08} (to name a few) in order to recognize the central and enduring status of hydrogen in physics.  Because of the well-developed theory and precise measurements \cite{Robilliard11,Eides01,Karshenboim05}, hydrogen is an attractive test bed for ruling out or setting bounds on new physics.  In this work we will extend the study of possible violations of Lorentz and CPT symmetry in hydrogen in the context of the Standard-Model Extension (SME) of Colladay and Kosteleck\'y \cite{Colladay97,Colladay98}.

The SME is a general framework for describing the effects that small violations of Lorentz and CPT symmetries would have on standard model processes.  It is based on the idea that Lorentz and CPT violation might arise through spontaneous symmetry violation of a more fundamental theory \cite{Kostelecky95}. The phenomenological effect of the SME is to add a number of new interactions to the standard model formed from the usual standard model fields coupled to spacetime constant background tensor fields.  These background fields are operationally just sets of symmetry-breaking parameters organized as scalars, vectors, and tensors of various ranks and symmetries.  The implications of the SME have been explored in many areas of physics.  A systematic overview organized as an extensive set of tables of bounds on the SME coefficients has been given by Kosteleck\'y and Russell \cite{Kostelecky11}.  Reviews of the SME have been given in Refs.~\cite{CPT,Bluhm06,Lehnert06}.

The SME causes small shifts in the energy levels of hydrogen that could, in principle, be detected by high precision spectroscopy.  In standard theory, the Dirac equation with a Coulomb potential is a convenient starting point for obtaining the energy levels of hydrogen--giving levels correctly through fine structure at the $O(m_e \alpha^4)$ level.  The SME leads to a modified Dirac equation, and through a non-relativistic reduction, to a modified effective Hamiltonian for use in a standard quantum mechanical framework.  The corrections to the effective Hamiltonian have the form of an expansion in powers of the relative momentum.  Since expectation values of the relative momentum $\vec p$ have a typical order $m_e \alpha$ (in units where $\hbar=c=1$) where $\alpha \approx 1/137.036$ is the fine structure constant, the low order terms in this expansion can be used to find the leading perturbations to the energy levels.  The leading SME corrections to hydrogen energy levels were worked out by Bluhm, Kosteleck\'y, and Russell \cite{Bluhm99} and Shore \cite{Shore05}, and additional leading order effects in hydrogen were considered by Ferreira and Moucherek \cite{Ferreira06} and Kharlanov and Zhukovsky \cite{Kharlanov07}.  By ``leading order'' we mean those corrections linear in SME parameters and of zeroth order in $\alpha$.  We will never consider terms of higher order than linear in the SME parameters since those should be completely negligible.  The ``higher order'' corrections that we obtain in this work are higher order in $\alpha$.  Results at order $\alpha^2$ times a subset of SME parameters have been reported by Bluhm {\it et al.} \cite{Bluhm99} and by Altschul \cite{Altschul10}; our results are more general.  Other Lorentz-violating corrections in hydrogen were analyzed by Belich {\it et al.} \cite{Belich06}.

In this work we obtain the corrections to the hydrogen energy levels at order $\alpha^2$ times the SME parameters for SME corrections to the electron, proton, and photon interactions in hydrogen.  We start with the SME-corrected Dirac equation and work out its non-relativistic expansion.  Our result agrees with that obtained earlier \cite{Kostelecky99a,Kostelecky99b,Lehnert04}.  We work out all corrections coming from terms linear in $\vec p$ (which vanish) and quadratic in $\vec p$ in the effective Hamiltonian.  The new corrections are at order $\alpha^2$ times SME parameters.  We obtain the complete set of such corrections in a compact form.  Next, we calculate the contributions at this order coming from SME corrections to the propagators of exchanged photons.  Finally, we apply our result to the $2S-1S$ transition in hydrogen, which has been measured with particularly high precision \cite{Parthey11}.

%%%%%%%%%%%%%%%%%%%%%%%%%%%%%%%%%%%%%%%%%%%%%%%%%%%

\section{Effective hamiltonian for low-energy SME corrections}

As a first step in obtaining the SME effects on non-relativistic systems, we find the extended Dirac equation for a fermion of mass $m$ in the presence of an electromagnetic field.  The appropriate SME Lagrangian has the form \cite{Colladay97,Kostelecky02a}
\be
{\cal L} = \frac{1}{2} i \bar \psi \Gamma^\nu \! \! \stackrel{\;\leftrightarrow}{D}_{\nu} \! \psi - \bar \psi M \psi
\ee
where $D_\nu = \partial_\nu + i q A_\nu$ is the covariant derivative, $F_{\mu \nu} = \partial_\mu A_\nu - \partial_\nu A_\mu$, and $\Gamma^\nu$ and $M$ contain the SME coefficients:
\bearray \label{SME_coefs}
\Gamma^\nu &=& \gamma^\nu + c^{\mu \nu} \gamma_\mu + d^{\mu \nu} \gamma_5 \gamma_\mu + e^\nu + i f^\nu \gamma_5 + \frac{1}{2} g^{\lambda \mu \nu} \sigma_{\lambda \mu} , \cr
M &=& m + a^\mu \gamma_\mu + b^\mu \gamma_5 \gamma_\mu + \frac{1}{2} H^{\mu \nu} \sigma_{\mu \nu}.
\eearray
All of the SME coefficients are real; $c_{\mu \nu}$ and $d_{\mu \nu}$ are traceless, $H_{\mu \nu}$ is antisymmetric, and $g_{\lambda \mu \nu}$ is antisymmetric on the $\lambda \mu$ indices.  Our gamma matrices are defined using the conventions of Itzykson and Zuber \cite{Itzykson80} and  we use a timelike metric.  While all of the SME parameters are Lorentz violating, the CPT-odd parameters $a^\mu$, $b^\mu$, $e^\mu$, $f^\mu$, and $g^{\lambda \mu \nu}$ also govern CPT violation \cite{Colladay98,Kostelecky02a}.
The Dirac equation that results from this Lagrangian is
\be\label{SME_dirac_eq}
\left ( i \Gamma^\mu D_\mu - M \right ) \psi = 0 .
\ee
The naive Hamiltonian associated with this Dirac equation is found by rearranging \eqref{SME_dirac_eq} into
\be
i \Gamma^0 \partial_0 \psi = \left ( - i \vec \Gamma \cdot \vec \nabla + q \Gamma^\mu A_\mu + M \right ) \psi,
\ee
or $i \partial_0 \psi = \tilde H \psi$ where
\be
\tilde H = \left ( \Gamma^0 \right )^{-1} \left ( -i \vec \Gamma \cdot \vec \nabla + q \Gamma^\mu A_\mu + M \right ) \psi .
\ee
However, $\tilde H$ is not hermitian and so is unacceptable.  As shown by \cite{Kostelecky99a,Kostelecky99b,Lehnert04}, we can repair the hermiticity problem by first making a spacetime-constant spinor field redefinition in the original Lagrangian:
\be
\psi = {\mathcal A} \chi ,
\ee
where ${\mathcal A}$ is chosen to restore the usual time derivative coupling.  In particular, we choose 
\be
{\mathcal A} = \left ( \gamma^0 \Gamma^0 \right )^{-1/2}.
\ee
This matrix is positive definite, invertible, and hermitian.  The modified Dirac equation takes the standard form $i \partial_0 \chi = H \chi$ where the hermitian Hamiltonian is
\be
H = {\mathcal A} \gamma^0 \vec \Gamma \cdot \vec \pi {\mathcal A} + {\mathcal A} \gamma^0 M {\mathcal A} + q A^0
\ee
with $\vec \pi = \vec p - q \vec A$.  We expand ${\mathcal A}$ to first order in the SME parameters and then apply the Foldy-Wouthuysen procedure \cite{Foldy50} to obtain the nonrelativistic expansion of the Hamiltonian.  We apply a succession of unitary transformations that act to diagonalize the Hamiltonian, using powers of $m$ to keep track of the order since each $1/m$ is associated with a factor of $\vec \pi$ and $\vec \pi$ is small in a non-relativistic system.  We specialize to the situation where $A^0$ is the Coulomb potential and $\vec A$ is a time-independent vector potential that leads to a constant magnetic field $\vec B$.  We work to first order in the magnetic field and neglect corrections to it involving either $\alpha$ or the SME parameters.  We have retained terms giving non-SME corrections of order $m \alpha^4$, the term of first order in $\vec B$, and all terms of $O(SME \cdot \alpha^0)$, $O(SME \cdot \alpha^1)$, and $O(SME \cdot \alpha^2)$.  The effective fermion Hamiltonian at this level of approximation is
\begin{eqnarray} \label{Heff}
H_{\rm eff} &=& \left ( m + \frac{p^2}{2m} - \frac{p^4}{8m^3} \right ) +q A_0 - \frac{q}{2m} \vec \sigma \cdot \vec B -\frac{q}{4m^2} \vec \sigma \cdot \left ( \vec E \times \vec p \right ) -\frac{q}{8 m^2} \vec \nabla \cdot \vec E + O(\alpha^5) \cr
&+& \left ( A + B_k \sigma^k \right ) + \left ( C_i + D_{i k} \sigma^k \right ) \frac{p^i}{m} + \left ( E_{i j} + F_{i j k} \sigma^k \right ) \frac{p^i p^j}{m^2} + O(SME\cdot\alpha^3)
\eearray
where
\bse \label{Coef_defs}
\bearray
A &=& a_0 - m e_0 - m c_{00} , \label{Adef} \\
B_k &=& -b_k + m d_{k 0} - \frac{1}{2} \epsilon_{k a b} \left ( m g_{a b 0} - H_{a b} \right ) , \label{Bdef} \\
C_i &=& a_i - m \left ( c_{i0}+c_{0i} \right ) - m e_i , \label{Cdef} \\
D_{i k} &=& -b_0 \delta_{i k} + \epsilon_{i k a} H_{a 0} + m \left ( d_{k i} + d_{00} \delta_{i k} \right ) - \frac{1}{2} m \epsilon_{a b k} g_{a b i} - m \epsilon_{i k a} g_{a 0 0} , \label{Ddef} \\
E_{i j} &=& -m c_{i j} - \frac{1}{2} m c_{0 0} \delta_{i j} , \label{Edef} \\
F_{i j k} &=& \delta_{j k} \tilde d_i + \frac{1}{2} \epsilon_{j k a} \left ( \epsilon_{a i p} b_p -2 m g_{a 0 i} - m g_{a i 0} \right ) , \label{Fdef}
\eearray
\ese
with $\tilde d_i = m d_{0 i}+\frac{1}{2} m d_{i 0} - \frac{1}{4} \epsilon_{i a b} H_{a b}$.  This effective Hamiltonian is a sum of the usual non-relativistic Hamiltonian $H_{NR}=m + p^2/(2m)+q A_0$, a magnetic $\vec \sigma \cdot \vec B$ interaction, fine structure contributions (relativistic kinetic energy, spin-orbit, Darwin), and the SME corrections.  We write $O(\alpha^5)$ and $O(SME \cdot \alpha^3)$ to indicate the orders of omitted terms when the fermion is bound in a Coulomb field.  While we distinguish between contravariant and covariant indices on SME tensors $a^\mu$, $b^\mu$, etc., we do not distinguish between superscripted and subscripted indices on purely three-dimensional objects like the Pauli matrices $\sigma_i=\sigma^i$, $\delta_{i j}$, $\epsilon_{i j k}$, and the coefficients $B_k$, $C_{i k}$, \dots, $F_{i j k}$.  We note that $a_0$ and $e_0$ enter only as a uniform shift to all levels, and $f^\mu$ drops out entirely to this order (as expected: see \cite{Colladay02,Altschul06,Fittante12}).  Our expression for the effective Hamiltonian is in agreement with the known result \cite{Kostelecky99a,Kostelecky99b,Lehnert04}.

%%%%%%%%%%%%%%%%%%%%%%%%%%%%%%%%%%%%%%%%%%%%%%%%%%%

\section{Fermion-sector energy shifts}

In this section we will work out the energy shifts on hydrogen states due to the SME effects on the electron and proton up to corrections of $O(SME \cdot \alpha^2)$.  We first describe the states in the absence of the SME.  The electron variables are labeled by $n$, $\ell$, $j$, $m_j$ where the total electron angular momentum $\vec J=\vec L + \vec S$ has a quantum number $j=\ell \pm 1/2$ in terms of the orbital angular momentum quantum number $\ell$.  The states $\vert n, \ell, j, m_j \rangle$ are approximate eigenstates of the Hamiltonian including fine structure and Lamb shift contributions.  The small corrections to these states that appear in the exact eigenstates ({\it i.e.} including fine structure and Lamb shift) do not affect SME corrections to $O(SME \cdot \alpha^2)$.  Explicit forms for the $\vert n, \ell, j, m_j \rangle$ states are
\be \label{j_states}
\vert n, \ell, j=\ell \pm 1/2, m_j \rangle = \pm c(\ell,\pm m_j) \vert n, \ell, m_j-1/2 \rangle \; \vert +1/2 \rangle_e + c(\ell,\mp m_j) \vert n, \ell, m_j+1/2 \rangle \; \vert -1/2 \rangle_e \, ,
\ee
where the $\vert n, \ell, m \rangle$ are the spinless Coulomb bound states with $\langle \vec r \, \vert n, \ell, m \rangle = R_{n \ell}(r) Y_\ell^m(\theta, \phi)$, the $\vert m_s \rangle_e$ are the electron spin states, and 
\be 
c(\ell, m)=\left ( \frac{\ell+m+1/2}{2 \ell+1} \right )^{1/2}.
\ee
More explicitly, one has
\be \label{wf_r}
\psi_{n, \ell, j=\ell \pm 1/2, m_j}(\vec r) = \langle \vec r \, \vert n, \ell, j=\ell \pm 1/2, m_j \rangle = R_{n \ell}(r) \begin{pmatrix} \pm c(\ell,\pm m_j) Y_\ell^{m_j-1/2}(\theta,\phi) \\ c(\ell,\mp m_j) Y_\ell^{m_j+1/2}(\theta,\phi) \end{pmatrix}.
\ee
We abbreviate these states as $\vert n, \ell, j=\ell \pm 1/2, m_j \rangle \equiv \vert m_j \rangle_\pm$.  Only the $+$ states exist when $\ell=0$.

The proton spin $\vec I$ affects the energy levels through the magnetic hyperfine interaction.  The total angular momentum $\vec F = \vec J + \vec I = (\vec L + \vec S)+\vec I$ has quantum number $f=j \pm' 1/2$.  It is conserved due to rotational symmetry, so the states having various values of $F_z$, with eigenvalues $\hbar m_f$, would be degenerate in the absence of an external magnetic field, but all other degeneracies are broken.  These states have the form
\bearray
\label{f_states} \vert n ,\ell, j=\ell \pm 1/2, f=j \pm' 1/2, m_f \rangle &=& \pm' c(j, \pm' m_f) \vert m_f-1/2 \; \rangle_\pm \; \vert +1/2 \rangle_p \cr
&\hbox{}& + c(j, \mp' m_f )  \vert m_f+1/2 \rangle_\pm \; \vert -1/2 \rangle_p \cr
&\equiv& \vert m_f \rangle_{\pm , \pm'} 
\eearray
where the $\vert m_i \rangle_p$ are the proton spin states.  We imagine our atom to be in the presence of a small external magnetic field so that the states $\vert m_f \rangle_{\pm, \pm'}$ are all non-degenerate.  This allows us to use the formalism of non-degenerate perturbation theory to calculation the SME corrections.  In section \ref{discussion} we will discuss the situation where the external magnetic field is not present.

Both the electron and the proton in the hydrogen atom are affected by SME corrections of the form given in \eqref{Heff}, so there should really be one set of SME corrections involving the electron variables (mass, momentum, spin matrices, and SME coefficients) and a second set of SME corrections involving the proton variables.  The electron and proton momenta sum to zero (in the center of mass frame), and have magnitude proportional to $\mu \alpha \approx m_e \alpha$ in a Coulombic bound state since the reduced mass $\mu \approx m_e$.  It would seem that proton corrections such as $p^i p^j/m_p^2$ would be negligible compared to electron ones from $p^i p^j/m_e^2$ given the small mass ratio $m_e/m_p \approx 1/1836$, but we have no {\it a prior\/i} knowledge of the size of the electron SME coefficients relative to the proton SME coefficients, so corrections of order $O(SME \cdot \alpha^2)$ for the proton might be comparable in size to those for the electron.  In any case, we have calculated both electron and proton corrections to order $O(SME \cdot \alpha^2)$.

We first consider the $O(SME \cdot \alpha^0)$ corrections that arise from terms in $H_{\rm eff}$ that are linear in the SME coefficients and independent of $\vec p$.  These terms are spatial constants so the matrix elements are easy to evaluate.  We can write these terms as
\be H_{\rm eff} \rightarrow A^e +A^p + B^e_k \sigma_e^k +B^p_k \sigma^k_p \ee
with $A^e$ and $A^p$ given by expressions like \eqref{Adef} and $B^e_k$ and $B^p_k$ by expressions like \eqref{Bdef}.  We use the Wigner-Ekhart theorem \cite{Edmonds57} to relate the $f=1$ operators $\vec \sigma_e$ and $\vec \sigma_p$ to $\vec F$ inside the expectation values, finding $\vec \sigma_e \rightarrow \xi_e \vec F$, $\vec \sigma_p \rightarrow \xi_p \vec F$ where the explicit values
\be \label{Wigner_Ekhart_results}
\xi_e = \pm \frac{2(2j+1\mp' 1)}{(2 \ell+1)(2j+1)} \; , \quad \xi_p = \pm' \frac{2}{2j+1}
\ee
were obtained by direct evaluation of the expectation values using the states \eqref{f_states}.  We find the $O(SME \cdot \alpha^0)$ energy shift to be
\be \label{E0_result}
E^0_{\rm SME} = \langle m_f \vert \left ( A^e + A^p + B^e_k \sigma_e^k + B^p_k \sigma_p^k \right ) \vert m_f \rangle_{\pm, \pm'} = \left ( A^e + A^p \right ) + \left ( \xi_e B^e_3+\xi_p B^p_3 \right ) m_f .
\ee
This $z$-axis is determined by the direction of the magnetic field and will be different from the $Z$-axis of the conventional sun-centered inertial frame \cite{Bluhm02,Bluhm03}.

The terms in $H_{\rm eff}$ that are linear in the momentum make contributions of $O(SME \cdot \alpha^1)$ to matrix elements, but do not contribute to expectation values.  One way to understand this is by thinking about parity: the states $\vert m_f \rangle_{\pm , \pm'}$ of \eqref{f_states} have definite orbital angular momentum $\ell$ and parity $(-1)^\ell$, and the matrix element of $\vec p$ (an odd-parity operator) between two states of the same parity must vanish.

The terms in $H_{\rm eff}$ that are quadratic in $\vec p$ make contributions of $O(SME \cdot \alpha^2)$.  These terms have the forms
\be \label{Heff_2}
H_{\rm eff} \rightarrow E^e_{i j} \frac{p^i p^j}{m_e^2} + F^e_{i j k} \frac{p^i p^j \sigma_e^k}{m_e^2} 
\ee
for the electron, with corresponding terms for the proton.  The energy corrections are given by the expectation value of \eqref{Heff_2} in the states \eqref{f_states}, so we must evaluate $\langle m_f \vert p^i p^j \sigma_e^\kappa \vert m_f \rangle_{\pm , \pm'}$ (where $\sigma_e^0 \equiv 1$) in these states.  We find it convenient to evaluate these expectation values in momentum space using wave functions
\be \label{wf_p}
\psi_{n, \ell, j=\ell \pm 1/2, m_j}(\vec p) = \tilde R_{n \ell}(p) \begin{pmatrix} \pm c(\ell,\pm m_j) Y_\ell^{m_j-1/2}(\theta,\phi) \\ c(\ell,\mp m_j) Y_\ell^{m_j+1/2}(\theta,\phi) \end{pmatrix}
\ee
instead of \eqref{wf_r}, where now $\tilde R_{n \ell}(p)$ is the momentum space radial function (given in terms of Gegenbauer polynomials, see \cite{Podolsky29,Fock36}) and $\theta$, $\phi$ represent the angles of $\vec p$.  The radial dependence of these expectation values factors out:
\be
\langle m_f \vert p^i p^j \sigma_e^\kappa \vert m_f \rangle_{\pm, \pm'} = \langle m_f \vert p^2 \hat p^i \hat p^j \sigma_e^\kappa \vert m_f  \rangle_{\pm, \pm'} = \langle p^2 \rangle_n \langle m_f \vert \hat p^i \hat p^j \sigma_e^\kappa \vert m_f \rangle_{\pm, \pm'} .
\ee
The expectation value of $p^2$ can be evaluated using the virial theorem in the usual way:
\be
\langle p^2 \rangle_n= -2m_e E_n = \frac{m_e^2 \alpha^2}{n^2}
\ee
where $E_n=-m_e \alpha^2/(2 n^2)$ are the Bohr energies.  We decompose $\hat p^i \hat p^j$ into irreducible tensors by writing
\be \label{irreducible}
\hat p^i \hat p^j = X_{i j} + \frac{1}{3} \delta_{i j}
\ee
where
\be \label{harmonics}
X_{i j} = \hat p^i \hat p^j - \frac{1}{3} \delta_{i j} = \sum_{\mu=-2}^2 C^\mu_{i j} Y_2^\mu(\theta,\phi) .
\ee
The coefficients $C^\mu_{i j}$ are given in Table \ref{coefficients}.

\begin{table}
\caption{\label{coefficients} Table of coefficients for writing $\hat p^i \hat p^j$ in terms of $Y^\mu_2(\theta, \phi)$.  The coefficient $C^\mu_{i j}=C^\mu_{j i}$ equals $\sqrt{2 \pi/15}$ times the number given at the appropriate spot in the table.}
\begin{ruledtabular}
\begin{tabular}{c | ccccc}
$i \, j$ & $\mu=-2$ & $\mu=-1$ & $\mu=0$ & $\mu=1$ & $\mu=2$  \\
\hline\noalign{\smallskip}
$1 \, 1$ & $1$ & $0$ & $-\sqrt{2/3}$ & $0$ & $1$ \\
$2 \, 2$ & $-1$ & $0$ & $-\sqrt{2/3}$ & $0$ & $-1$ \\
$3 \, 3$ & $0$ & $0$ & $2 \sqrt{2/3}$ & $0$ & $0$ \\
$1 \, 2$ & $i$ & $0$ & $0$ & $0$ & $-i$ \\
$1 \, 3$ & $0$ & $1$ & $0$ & $-1$ & $0$ \\
$2 \, 3$ & $0$ & $i$ & $0$ & $i$ & $0$ \\
\end{tabular}
\end{ruledtabular}
\end{table}

We build up to the evaluation of $\langle m_f \vert p^i p^j \sigma_e^\kappa \vert m_f \rangle_{\pm, \pm'}$ in three steps.  First, we obtain the matrix elements of $p^i p^j$ in the spinless states $\vert n, \ell, m \rangle$.  We use these to find the expectation values $\langle m_j \vert p^i p^j \sigma_e^\kappa \vert m_j \rangle_\pm$ in the states $\vert m_j \rangle_\pm$ of the electron with spin.  Finally we evaluate the expectation values $\langle m_f \vert p^i p^j \sigma_e^\kappa \vert m_f \rangle_{\pm, \pm'}$.

The spinless matrix elements are easy to obtain by use of the decomposition of $\hat p^i \hat p^j$ described in 
\eqref{irreducible} and \eqref{harmonics}.  One has
\bearray
\langle n, \ell, m \vert \hat p^i \hat p^j \vert n, \ell, m' \rangle &=& \langle n, \ell, m \vert \left \{ \sum_{\mu=-2}^2 C^\mu_{i j} Y^\mu_2(\theta, \phi) + \frac{1}{3} \delta_{i j} \right \} \vert n, \ell, m' \rangle \cr
&=& C^{m-m'}_{i j} I_{\ell, m, m'} + \frac{1}{3} \delta_{i j} \delta_{m m'}
\eearray
where the integral over solid angle
\be
I_{\ell, m, m'} = \int d \Omega \, Y^{m*}_\ell(\theta, \phi) Y^{m-m'}_2 (\theta, \phi) Y^{m'}_\ell(\theta, \phi)
=\sqrt{\frac{5}{4 \pi}} C^{\ell, m}_{\ell, m'; 2, m-m'} C^{\ell,0}_{\ell,0; 2,0}
\ee
is given in terms of Clebsch-Gordon coefficients \cite{Edmonds57}.  Matrix elements of $p^i p^j$ are obtained from those of $\hat p^i \hat p^j$ by multiplication by $\langle p^2 \rangle_n$.

When electron spin is included we need the matrix elements
\be
\langle m_j \vert \hat p^i \hat p^j \sigma_e^\kappa \vert m_j' \rangle_\pm = \int d \Omega \, \left ( \pm c(\ell,\pm m_j) Y_\ell^{m_j-1/2}(\theta,\phi) , c(\ell,\mp m_j) Y_\ell^{m_j+1/2}(\theta,\phi) \right )^* \hat p^i \hat p^j \sigma_e^\kappa  \begin{pmatrix} \pm c(\ell,\pm m_j') Y_\ell^{m_j'-1/2}(\theta,\phi) \\ c(\ell,\mp m_j') Y_\ell^{m_j'+1/2}(\theta,\phi) \end{pmatrix}.
\ee
The $2 \times 2$ matrix algebra gives, for example,
\bearray
\langle m_j \vert \hat p^i \hat p^j \sigma_e^0 \vert m_j' \rangle_\pm &=& c(\ell, \pm m_j) c(\ell, \pm m_j') \langle n, \ell, m_j-1/2 \vert \hat p^i \hat p^j \vert n, \ell, m_j'-1/2 \rangle \cr &+& c(\ell, \mp m_j) c(\ell, \mp m_j') \langle n, \ell, m_j+1/2 \vert \hat p^i \hat p^j \vert n, \ell, m_j'+1/2 \rangle ,
\eearray
with analogous expressions when $\kappa$ is 1, 2, or 3.  It turns out that we need diagonal matrix elements only.  The results are:
\bse
\bearray
\langle m_j \vert \hat p^i \hat p^j \vert m_j \rangle_\pm &=& \left ( \frac{j(j+1)+m_j^2}{2j(j+1)} \right ) \frac{ \delta_{i 1} \delta_{j 1} + \delta_{i 2} \delta_{j 2}}{2} + \left ( \frac{\left ( j(j+1)-m_j^2 \right )}{2j(j+1)} \right ) \delta_{i 3} \delta_{j 3} , \\
\langle m_j \vert \hat p^i \hat p^j \sigma_e^1 \vert m_j \rangle_\pm &=& \mp m_j \left ( \frac{ (2 \ell+1)^2-4 m_j^2}{2(2\ell+3)(2 \ell+1)(2 \ell-1)} \right ) \left ( \delta_{i 1} \delta_{j 3} + \delta_{i 3} \delta_{j 1} \right ) , \\
\langle m_j \vert \hat p^i \hat p^j \sigma_e^2 \vert m_j \rangle_\pm &=& \mp m_j \left ( \frac{ (2 \ell+1)^2-4 m_j^2}{2(2\ell+3)(2 \ell+1)(2 \ell-1)} \right ) \left ( \delta_{i 2} \delta_{j 3} + \delta_{i 3} \delta_{j 2} \right ) , \\ \nonumber
\langle m_j \vert \hat p^i \hat p^j \sigma_e^3 \vert m_j \rangle_\pm &=& \pm m_j \left ( \frac {(2 \ell+3)(2 \ell-1) \mp 2(2 \ell+1)+4 m_j^2}{(2 \ell+3)(2\ell+1)(2\ell-1)} \right ) \frac{ \delta_{i 1} \delta_{j 1} + \delta_{i 2} \delta_{j 2}}{2} \\
&\pm& m_j \left ( \frac{(2 \ell+3)(2 \ell-1) \pm 2(2\ell+1)-4 m_j^2}{(2\ell+3)(2\ell+1)(2 \ell-1)} \right ) \delta_{i 3} \delta_{j 3} .
\eearray
\ese

Finally, we are in a position to obtain the full expectation values $\langle m_f \vert p^i p^j \sigma_e^\kappa \vert m_f \rangle_{\pm, \pm'}$.  We find these expectation values by using the states \eqref{f_states} and noticing that the operator $p^i p^j \sigma_e^\kappa$ is independent of proton spin, so that
\be
\langle m_f \vert p^i p^j \sigma_e^\kappa \vert m_f \rangle_{\pm, \pm'} = c^2(j,\pm' m_f) \langle m_f-1/2 \vert p^i p^j \sigma_e^\kappa \vert m_f-1/2 \rangle_\pm + c^2(j,\mp' m_f) \langle m_f+1/2 \vert p^i p^j \sigma_e^\kappa \vert m_f+1/2 \rangle_\pm .
\ee
Our results for the required expectation values are:
\bse \label{f_expec}
\bearray
\langle m_f \vert \hat p^i \hat p^j \vert m_f \rangle_{\pm, \pm'} &=& \left ( \frac{1}{2} + \frac{1+4 m_f^2}{8j(j+1)} \mp' \frac{m_f^2}{j(j+1)(2j+1)} \right ) \frac{\delta_{i 1} \delta_{j 1} + \delta_{i 2} \delta_{j 2}}{2} \nonumber \label{f_expec_a} \\
&+& \left ( \frac{1}{2}- \frac{1+4 m_f^2}{8j(j+1)} \pm'  \frac{m_f^2}{j(j+1)(2j+1)} \right ) \delta_{i 3} \delta_{j 3} \\
\langle m_f \vert \hat p^i \hat p^j \sigma_e^1 \vert m_f \rangle_{\pm, \pm'} &=& \frac{\mp 2 m_f}{(2 \ell+3)(2 \ell+1)(2 \ell-1)} \nonumber \\
&\times& \left ( \ell(\ell+1)-1/2-m_f^2 \mp' \frac{\ell(\ell+1)-3 m_f^2}{2j+1} \right ) \left ( \delta_{i 1} \delta_{j 3} + \delta_{i 3} \delta_{j 1} \right ) , \\
\langle m_f \vert \hat p^i \hat p^j \sigma_e^2 \vert m_f \rangle_{\pm, \pm'} &=& \frac{\mp 2 m_f}{(2 \ell+3)(2 \ell+1)(2 \ell-1)} \nonumber \\
&\times& \left ( \ell(\ell+1)-1/2-m_f^2 \mp' \frac{\ell(\ell+1)-3 m_f^2}{2j+1} \right ) \left ( \delta_{i 2} \delta_{j 3} + \delta_{i 3} \delta_{j 2} \right ) , \\
\langle m_f \vert \hat p^i \hat p^j \sigma_e^3 \vert m_f \rangle_{\pm, \pm'} &=& \frac{\pm 4 m_f}{(2 \ell+3)(2 \ell+1)(2 \ell-1)}
\nonumber \\
&\times& \Bigg [ \left (  \ell(\ell+1) \mp (\ell+1/2) + m_f^2 \mp' \frac{\ell(\ell+1)-1/2 \mp (\ell+1/2) + 3 m_f^2}{2j+1} \right )
\frac{ \delta_{i 1} \delta_{j 1} + \delta_{i 2} \delta_{j 2} }{2} \nonumber \\
&+& \left (  \ell(\ell+1) - 3/2 \pm (\ell+1/2) - m_f^2  \mp' \frac{\ell(\ell+1)-1 \pm (\ell+1/2) - 3 m_f^2}{2j+1} \right ) \delta_{i 3} \delta_{j 3} \Bigg ] .
\eearray
\ese
Some consequences of \eqref{f_expec} are
\bse \label{cons}
\bearray 
\langle m_f \vert \hat p \cdot \hat p \vert m_f \rangle_{\pm, \pm'} &=& \langle m_f  \vert m_f \rangle_{\pm, \pm'} = 1 ,  \label{cons_a} \\
\langle m_f \vert \hat p \cdot \hat p \sigma_e^k \vert m_f \rangle_{\pm, \pm'} &=& \langle m_f \vert \sigma_e^k \vert m_f \rangle_{\pm, \pm'} = \pm \frac{ 2 m_f (2j+1\mp' 1)}{(2\ell+1)(2 j+1)} \delta_{k 3} , \label{cons_b} \\
\langle m_f \vert \hat p^i \hat p \cdot \vec \sigma_e \vert m_f \rangle_{\pm, \pm'} &=& \langle m_f \vert \hat p^i \hat p^j \sigma_e^k \vert m_f \rangle_{\pm, \pm'}  \delta_{j k} = \frac{m_f (2j+1\mp' 1)}{2j(j+1) (2j+1)} \delta_{i 3} = \frac{2 m_f}{(2f+1)(2j+1)} \delta_{i 3} , \label{cons_c} \\
\langle m_f \vert \hat p^i \left ( \hat p \times \vec \sigma_e \right )^a \vert m_f \rangle_{\pm, \pm'} &=& \langle m_f \vert \hat p^i \hat p^j \sigma_e^k \vert m_f \rangle_{\pm, \pm'} \epsilon_{j k a} = \mp \frac{m_f (2j+1\mp' 1)}{4j(j+1)} \epsilon_{i a 3} = \mp \frac{ m_f}{2f+1} \epsilon_{i a 3} , \label{cons_d}
\eearray
\ese
where \eqref{cons_a} and \eqref{cons_b} serve as checks, and \eqref{cons_c} and \eqref{cons_d} will be needed below.  In the reductions of \eqref{cons_c} and \eqref{cons_d} we have applied the useful identity $(2j+1 \mp' 1)=4j(j+1)/(2f+1)$, and we record here the additional identity $(2j+1 \mp' 2) = (2j+3)(2j+1)(2j-1)/(4f(f+1))$.  As always, expectation values of $p^i p^j \sigma_e^\kappa$ are obtained from those of $\hat p^i \hat p^j \sigma_e^\kappa$ by multiplication by $\langle p^2 \rangle_n = (m_e \alpha/n)^2$.

The corresponding proton expectation values of $\hat p^i \hat p^j \sigma^k_p$ are obtained by calculations similar to those used for $\hat p^i \hat p^j \sigma^k_e$.  We start from \eqref{f_states} and find
\bse \label{f_expec_pre_p}
\bearray 
\langle m_f \vert \hat p^i \hat p^j \sigma^1_p \vert m_f \rangle_{\pm, \pm'} &=& \pm' 2 c(j,\pm' m_f) c(j,\mp' m_f) Re \left [ \langle m_f-1/2 \vert \hat p^i \hat p^j \vert m_f+1/2 \rangle_\pm \right ] , \\
\langle m_f \vert \hat p^i \hat p^j \sigma^2_p \vert m_f \rangle_{\pm, \pm'} &=& \pm' 2 c(j,\pm' m_f) c(j,\mp' m_f) Im \left [ \langle m_f-1/2 \vert \hat p^i \hat p^j \vert m_f+1/2 \rangle_\pm \right ] , \\
\langle m_f \vert \hat p^i \hat p^j \sigma^3_p \vert m_f \rangle_{\pm, \pm'} &=& c^2(j,\pm' m_f) \langle m_f-1/2 \vert \hat p^i \hat p^j \vert m_f-1/2 \rangle_\pm - c^2(j,\mp' m_f) \langle m_f+1/2 \vert \hat p^i \hat p^j \vert m_f+1/2 \rangle_\pm .
\eearray
\ese
The required matrix elements in the orbital plus spin states $\vert n, \ell, j=\ell \pm 1/2, m_j \rangle$ of \eqref{j_states} are
\bse \label{j_matrix_elements}
\bearray
\langle m_f-1/2 \vert \hat p^i \hat p^j \vert m_f+1/2 \rangle_\pm &=& \frac{-m_f \sqrt{(j+1/2)^2-m_f^2}}{4j(j+1)} \left \{ \left (\delta_{i 1} \delta_{j 3} + \delta_{i 3} \delta_{j 1} \right ) + i \left ( \delta_{i 2} \delta_{j 3}+\delta_{i 3} \delta_{j 2} \right ) \right \} , \\
\langle m_f-1/2 \vert \hat p^i \hat p^j \vert m_f-1/2 \rangle_\pm &=& \left ( \frac{(j+1/2)^2+m_f^2-m_f}{2j(j+1)} \right ) \frac{ \delta_{i 1} \delta_{j 1} + \delta_{i 1} \delta_{j 2} }{2} \nonumber \\ &+& \left ( \frac{(j+1/2)^2-1/2-m_f^2+ m_f}{2j(j+1)} \right ) \delta_{i 3} \delta_{j 3} , \\
\langle m_f+1/2 \vert \hat p^i \hat p^j \vert m_f+1/2 \rangle_\pm &=& \left ( \frac{(j+1/2)^2+m_f^2+m_f}{2j(j+1)} \right ) \frac{ \delta_{i 1} \delta_{j 1} + \delta_{i 2} \delta_{j 2} }{2} \nonumber \\ &+& \left ( \frac{(j+1/2)^2-1/2-m_f^2-m_f}{2j(j+1)} \right ) \delta_{i 3} \delta_{j 3} .
\eearray
\ese
These lead, through equations \eqref{f_expec_pre_p}, to
\bse \label{f_expec_p}
\bearray
\langle m_f \vert \hat p^i \hat p^j \sigma^1_p \vert m_f \rangle_{\pm, \pm'} &=& \frac{\mp' m_f \left ( (j+1/2)^2-m_f^2 \right )}{2j(j+1)(2j+1)} \left ( \delta_{i 1} \delta_{j 3} + \delta_{i 3} \delta_{j 1} \right ) , \\
\langle m_f \vert \hat p^i \hat p^j \sigma^2_p \vert m_f \rangle_{\pm, \pm'} &=& \frac{\mp' m_f \left ( (j+1/2)^2-m_f^2 \right )}{2j(j+1)(2j+1)} \left ( \delta_{i 2} \delta_{j 3} + \delta_{i 3} \delta_{j 2} \right ) , \\
\langle m_f \vert \hat p^i \hat p^j \sigma^3_p \vert m_f \rangle_{\pm, \pm'} &=& \frac{\pm' m_f}{j(j+1)(2j+1)} 
\Bigg [ \Big ( (j+1/2)^2+m_f^2 \mp' (j+1/2) \Big ) \left ( \frac{ \delta_{i 1} \delta_{j 1} + \delta_{i 2} \delta_{j 2} }{2} \right ) \nonumber \\
&+& \left ( (j+1/2)^2-\frac{1}{2} - m_f^2 \pm' (j+1/2) \right ) \delta_{i 3} \delta_{j 3} \Bigg ] .
\eearray
\ese
Consequences of \eqref{f_expec_p} include
\bse
\bearray
\langle m_f \vert \hat p \cdot \hat p \sigma_p^k \vert m_f \rangle_{\pm, \pm'} &=& \langle m_f \vert \sigma^k_p \vert m_f \rangle_{\pm, \pm'} = \pm' \frac{ 2 m_f }{2 j+1} \delta_{k 3} , \label{consp_a} \\
\langle m_f \vert \hat p^i \hat p \cdot \vec \sigma_p \vert m_f \rangle_{\pm, \pm'} &=&  \frac{2 m_f}{(2f+1)(2j+1)} \delta_{i 3} , \label{consp_b} \\
\langle m_f \vert \hat p^i \left ( \hat p \times \vec \sigma_p \right )^a \vert m_f \rangle_{\pm, \pm'} &=& \mp' \frac{ m_f}{2f+1} \epsilon_{i a 3} . \label{consp_c}
\eearray
\ese
Expression \eqref{consp_a} is a check showing consistency with \eqref{Wigner_Ekhart_results}, while \eqref{consp_b} and \eqref{consp_c} will be needed for the energy shift calculations.  We note the similarity (and identity when $\pm = \pm'$) between the electron spin results of \eqref{cons} and the proton spin results here.  

The $O(SME \cdot \alpha^2)$ energy level shifts coming from SME electron and proton interactions for a state with quantum numbers $n$, $\ell$,  $j=\ell \pm 1/2$,  $f=j \pm' 1/2$, and $m_f$ are
\be \label{fermion_result1}
E^{2ep}_{\rm SME} = \Big \{ \langle m_f \vert \hat p^i \hat p^j \vert m_f \rangle_{\pm, \pm'} \left ( E^e_{i j} + \epsilon E^p_{i j} \right )
+ \langle m_f \vert \hat p^i \hat p^j \left (\sigma_e^k F^e_{i j k} + \sigma_p^k \epsilon F^p_{i j k} \right ) \vert m_f \rangle_{\pm, \pm'} \Big \} \left ( \frac{\alpha}{n} \right )^2 ,
\ee
where $\epsilon = (m_e/m_p)^2$.  More explicitly, the $O(SME \cdot \alpha^2)$ energy corrections from these sources are:
\bearray \label{fermion_result}
E^{2ep}_{\rm SME} &=& \Bigg \{ -\frac{5}{6} \left (\tilde c^e_{0 0}+\epsilon \tilde c^p_{0 0} \right ) + \frac{(2j+3)(2j-1)}{16j(j+1)} \left ( \frac{1}{3} - \frac{m_f^2}{f(f+1)} \right ) \left (\tilde c^e_Q + \epsilon \tilde c^p_Q \right ) \nonumber \\
&+& \frac{m_f}{2f+1} \left ( \frac{2}{2j+1} \left (\tilde d^e_3 + \epsilon \tilde d^p_3 \right ) + \left (\mp \tilde g^e_{D 3} \mp' \epsilon \tilde g^p_{D 3} \right )\right ) \Bigg \}  \left ( \frac{\alpha}{n} \right )^2 .
\eearray
We find that only four combinations of parameters enter into the final result: $\tilde c_{0 0} = m c_{0 0}$ (where $c_{0 0}=c_{i i}$ because $c_{\mu \nu}$ is traceless), $\tilde c_Q \equiv m \left ( c_{1 1}+c_{2 2} - 2 c_{3 3} \right )$, $\tilde d_3=m d_{0 3}+\frac{1}{2} m d_{3 0} - \frac{1}{2} H_{1 2}$, and $\tilde g_{D 3} \equiv -b_3+m (g_{1 0 2} - g_{2 0 1} + g_{1 2 0} )$.  (The definitions were adapted from Table XVIII of Ref.~\cite{Kostelecky11}.)

%%%%%%%%%%%%%%%%%%%%%%%%%%%%%%%%%%%%%%%%%%%%%%%%%%%

\section{Photon-sector energy shifts}

In this section we calculate the energy level corrections at $O(SME \cdot \alpha^2)$ coming from SME corrections to the propagators for exchange photons.  The graphs in question are shown in Fig.~\ref{fig1}.  The Feynman rules for SME corrections to the photon propagator are given, {\it e.g.\/}, in the appendix to \cite{Kostelecky02a}.  There are two corrections: one (indicated by a large filled dot), has the form $-2i q^\alpha q^\beta (k_F)_{\alpha \mu \beta \nu}$, where $\mu$ and $\nu$ are photon indices and momentum $q$ travels from the $\mu$ to the $\nu$ index.  The second (indicated by a cross) has the form $2 (k_{AF})^\alpha \epsilon_{\alpha \mu \beta \nu} q^\beta$, where $\epsilon_{\alpha \mu \beta \nu}$ is the four-dimensional Levi-Civita symbol ($\epsilon_{0 1 2 3}=-1$).  The counting rules for QED bound states such as hydrogen allow us to estimate the orders in $\alpha$ of the graphs shown.  Each interaction vertex brings a factor of $e \propto \alpha^{1/2}$.  Three-momenta are of order $\alpha$ and energies are $\alpha^2$.  Wave functions scale as $\alpha^{-3/2}$ in momentum space.  The first (Coulomb interaction) diagram has an interaction factor $(e)(1/\vec q\, ^2)(-e)=-4 \pi \alpha/\vec q\, ^2$ and a corresponding energy given by
\be
E_C = \int \frac{d^3 p}{(2 \pi)^3} \frac{d^3 p'}{(2 \pi)^3} \langle \vec p \, \vert m_f \rangle^*_{\pm, \pm'}  \left ( \frac{-4 \pi \alpha}{\vec q\, ^2} \right ) \langle \vec p\, '  \vert m_f \rangle_{\pm, \pm'}
\ee
where $\vec q = \vec p - \vec p\, '$.  The scaling laws imply an order $O(\alpha^2)$ for this energy, and indeed the Fourier transform of the interaction is $-\alpha/r$ with expectation value $E_C = \langle -\alpha/r \rangle = 2E_n = -m_e (\alpha/n)^2$.  We use Coulomb gauge for the photon propagator, and take the QED Feynman rules for this non-relativistic system from Non-Relativisitc QED (NRQED), whose rules have been tabulated by, for example, Kinoshita and Nio \cite{Kinoshita96} or Labelle \cite{Labelle98}.
%%%%%%
\begin{figure}
\includegraphics{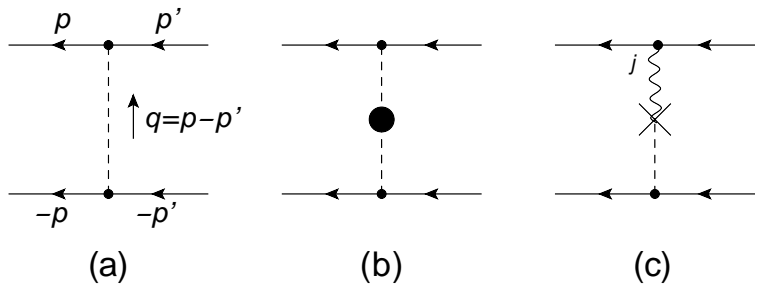}
\caption{\label{fig1} Graphs involving photon exchange between the electron (top) and proton (bottom).  Part (a) shows Coulomb exchange photon that is responsible for the dominant binding.  Parts (b) and (c) show SME corrections.  In (b) the interaction involves the SME coefficients $(k_F)_{\alpha \mu \beta \nu}$ acting between two Coulomb photons.  In (c) the SME correction with coefficients $(k_{AF})^\alpha$ acts between a Coulomb photon (connected to the proton) and a transverse photon (attached to the electron).}
\end{figure}
%%%%%%

The graph of Fig~1(b) has, compared to the Coulomb contribution of Fig.~1(a), two extra powers of the photon three-momentum in the SME vertex and an extra Coulomb propagator.  It thus has the same order, $\alpha^2$, as the Coulomb energy, but involves as well the SME parameters $(k_F)_{\alpha \mu \beta \nu}$.  Spatial parts of the Coulomb-gauge propagator do not contribute at this order in $\alpha$ because they would come with a ``dipole'' vertex $-e(\vec p\, '+\vec p\, )/(2 m)$ containing an extra power of the three-momenta.  The interaction factor arising from the graph of Fig.~1(b) is $(e)(1/\vec q\, ^2) i \left ( -2i q^i q^j (k_F)_{i 0 j 0} \right )(1/\vec q\, ^2)(-e) = -2e^2 (k_F)_{i 0 j 0} q^i q^j/\vec q\, ^4$ where we have used the antisymmetry of $(k_F)_{\alpha \mu \beta \nu}$ in the first and in the second pair of indices. \cite{Colladay98,Kostelecky02a}  The extra factor of $i$ before the SME Feynman rule factor is to convert to the conventions used by Refs.~\cite{Kinoshita96,Labelle98}.  The Fourier transform of the momentum-dependent part of the interaction term, $q^i q^j/\vec q\, ^4$, is $\left ( \delta_{i j} - \hat x^i \hat x^j \right )/(8 \pi r)$, as shown for example in Ref.~\cite{Berestetskii}, so the correction to the potential energy is $(\alpha / r) (k_F)_{i 0 j 0} \left ( \hat x^i \hat x^j - \delta_{i j} \right )$.  This is consistent with the potential found in Refs.~\cite{Kostelecky02b,Bailey04,Bailey10} from the solution of the SME--modified Maxwell equations. The corresponding correction to the energy is
\be
E_F = \int d^3 x \langle \vec x \, \vert m_f \rangle^*_{\pm, \pm'}  \left ( \frac{\alpha}{r} (k_F)_{i 0 j 0} \left ( \hat x^i \hat x^j - \delta_{i j} \right ) \right ) \langle \vec x \, \vert m_f \rangle_{\pm, \pm'}.
\ee
As in our earlier momentum space work, the radial part factorizes, giving $\langle \alpha / r \rangle_n = m_e (\alpha / n)^2$.  We are left with an angular expectation value
\be \label{E_F_result1}
E_F = m_e (k_F)_{i 0 j 0} \langle m_f \vert \left ( \hat x^i \hat x^j-\delta_{i j} \right ) \vert m_f \rangle_{\pm, \pm'} \left ( \frac{\alpha}{n} \right )^2
\ee
The angular expectation value here is the same in coordinate space as in momentum space, so we can use \eqref{f_expec_a} for the expectation value of $\hat x^i \hat x^j$.  We find it convenient to write this expectation value as
\be
\langle m_f \vert \hat x^i \hat x^j \vert m_f \rangle_{\pm, \pm'} = \langle m_f \vert \hat p^i \hat p^j \vert m_f \rangle_{\pm, \pm'} = \frac{1}{3} \delta_{i j} - \frac{(2j+3)(2j-1)}{16j(j+1)} \left ( \frac{1}{3} - \frac{m_f^2}{f(f+1)} \right ) \delta^q_{i j}
\ee
where $\delta^q_{i j} \equiv \delta_{i 1} \delta_{j 1} + \delta_{i 2} \delta_{j 2} - 2 \delta_{i 3} \delta_{j 3}$.  The resulting expression for the energy shift is
\be \label{E_F}
E_F = \left \{ -\frac{2}{3} \tilde \kappa_0  - \frac{(2j+3)(2j-1)}{16j(j+1)} \left ( \frac{1}{3} - \frac{m_f^2}{f(f+1)} \right ) \tilde \kappa_Q \right \} \left ( \frac{\alpha}{n} \right )^2 ,
\ee
where $\tilde \kappa_0 \equiv m_e \left ( (k_F)_{1 0 1 0} + (k_F)_{2 0 2 0} + (k_F)_{3 0 3 0} \right )$ and $\tilde \kappa_Q \equiv m_e \left ( (k_F)_{1 0 1 0} + (k_F)_{2 0 2 0} - 2 (k_F)_{3 0 3 0} \right )$.

The graph of Fig.~1(c) involves the $k_{AF}$ SME parameters.  Because of the Levi-Civita symbol in the Feynman rule, only one of the photons can be Coulomb, the other must be transverse.  We calculate the graph having the transverse photon attached to the electron because it is larger by a factor of $m_e/m_p$ than the one with the transverse photon connected to the proton.  The interaction factor for this graph is $\left ( -e (p+p')^a /(2 m_e) \right ) \left ( -\delta_{a k}/\vec q\, ^2 \right ) i \left ( 2(k_{AF})^i \epsilon_{i 0 j k} q^j \right ) \left (1/ \vec q\, ^2 \right ) (-e) = \left ( i e^2/m_e \right ) (k_{AF})^i \epsilon_{0 i j k} q^j (p+p')^k/\vec q\, ^4$.  We can write $(p+p')^k = (q + 2 p')^k \rightarrow 2 p'^k$, so the interaction factor becomes $\left ( -2i e^2/m_e \right ) (k_{AF})^i \epsilon_{i j k} q^j p'^k/\vec q \, ^4$.  This interaction factor is not just a function of the relative momentum $\vec q$, so we work out the Fourier analysis in more detail.  The energy shift from Fig.~1(c) is
\bearray
E_{AF} &=& \frac{-2 i e^2}{m_e} (k_{AF})^i \epsilon_{i j k} \int \frac{d^3 p}{(2 \pi)^3} \frac{d^3 p'}{(2 \pi)^3} \langle \vec p \, \vert m_f \rangle^*_{\pm, \pm'} \left ( \frac{q^j p'^k}{\vec q \, ^4} \right ) \langle \vec p\, '  \vert m_f \rangle_{\pm, \pm'} \cr
&=& \frac{-2 i e^2}{m_e} (k_{AF})^i \epsilon_{i j k} \int \frac{d^3 p}{(2 \pi)^3} \frac{d^3 p'}{(2 \pi)^3} d^3 x \, d^3 y \, \langle \vec x \, \vert m_f \rangle^*_{\pm, \pm'} \, e^{i \vec p \cdot \vec x} \left ( \frac{q^j}{\vec q \, ^4} \right ) i \nabla_y^k e^{-i \vec p\, ' \cdot \vec y} \langle \vec y\, \vert m_f \rangle_{\pm, \pm'} \cr
&=& \frac{-2 i e^2}{m_e} (k_{AF})^i \epsilon_{i j k} \int d^3 x \, d^3 y \frac{d^3 q}{(2 \pi)^3} \frac{d^3 p'}{(2 \pi)^3} \, \langle \vec x \, \vert m_f \rangle^*_{\pm, \pm'} \, e^{i \vec q \cdot \vec x} \left ( \frac{q^j}{\vec q \, ^4} \right ) e^{-i \vec p\, ' \cdot (\vec y-\vec x)} \left (-i \nabla_y^k \right ) \langle \vec y\, \vert m_f \rangle_{\pm, \pm'} \cr
&=& \frac{-2 i e^2}{m_e} (k_{AF})^i \epsilon_{i j k} \int d^3 x \, \langle \vec x \, \vert m_f \rangle^*_{\pm, \pm'} \left ( \frac{i x^j}{8 \pi r} \right ) \left (-i \nabla_x^k \right ) \langle \vec x\, \vert m_f \rangle_{\pm, \pm'}.
\eearray
So this energy is just proportional to the expectation value of the orbital angular momentum $\vec L = \vec x \times \vec p$.  We use the Wigner-Ekhart theorem as in \eqref{Wigner_Ekhart_results} to write $\vec L \rightarrow \xi_\ell \vec F$ inside the $\vert m_f \rangle_{\pm, \pm'}$ expectation values where
\be
\xi_\ell = \frac{16 \ell(\ell+1) j(j+1)}{(2 \ell+1)(2j+1)^2 (2f+1)} .
\ee
We found $\xi_\ell$ by direct evaluation using the states $\vert m_f \rangle_{\pm, \pm'}$ of \eqref{f_states}, and checked using $\xi_\ell + \xi_e/2 + \xi_p/2=1$, which follows from $\vec L + \vec \sigma_e/2 + \vec \sigma_p/2 = \vec F$.  The $E_{AF}$ energy shift is thus
\be \label{E_AF}
E_{AF} = \frac{1}{m_e} \left \langle \frac{\alpha}{r} \right \rangle \langle m_f \vert (k_{AF})^i L^i \vert m_f \rangle_{\pm, \pm'} = (k_{AF})^3 \xi_\ell m_f \left ( \frac{\alpha}{n} \right )^2.
\ee

%%%%%%%%%%%%%%%%%%%%%%%%%%%%%%%%%%%%%%%%%%%%%%%%%%%

\section{Result, and Application to the $2S-1S$ transition in hydrogen}

The total energy shift at order $O(SME \cdot \alpha^2)$ is the sum of the fermion-sector shift $E^{2ep}_{SME}$ of \eqref{fermion_result} and the two photon-sector contributions $E_F$ of \eqref{E_F} and $E_{AF}$ of \eqref{E_AF}.  It is
\bearray \label{E2_result}
E^2_{\rm SME} &=& \Bigg \{ -\frac{5}{6} \left (\tilde c^e_{0 0}+\epsilon \tilde c^p_{0 0} \right ) - \frac{2}{3} \tilde \kappa_0 + \frac{(2j+3)(2j-1)}{16j(j+1)} \left ( \frac{1}{3} - \frac{m_f^2}{f(f+1)} \right ) \left (\tilde c^e_Q + \epsilon \tilde c^p_Q - \tilde \kappa_Q \right ) \nonumber \\
&+& \frac{m_f}{2f+1} \left ( \frac{2}{2j+1} \left (\tilde d^e_3 + \epsilon \tilde d^p_3 \right ) + \left (\mp \tilde g^e_{D 3} \mp' \epsilon \tilde g^p_{D 3} \right )\right ) + m_f \xi_\ell (k_{AF})^3 \Bigg \}  \left ( \frac{\alpha}{n} \right )^2 .
\eearray
The SME coefficients appearing here are those for the lab frame.  The SME coefficients in the lab are related to those in the conventional sun-centered inertial frame by a Lorentz transformations involving both a boost and a rotation.  The expression for $E^2_{\rm SME}$ given by \eqref{E2_result} is our main result.

The transition between the $2S$ and $1S$ states is of particular interest because it has an anomalously small intrinsic broadness and it has been measured with great precision.  The $2S \rightarrow 1S$ electric dipole decay into a single photon is forbidden by the usual $\Delta \ell=\pm 1$ selection rule.  The decay happens by a slower two-photon transition with a lifetime of about $0.12 s$ (as opposed to the $1.6 ns$ lifetime of the $2P$ state). \cite{Breit40,Klarsfeld69}  The corresponding uncertainty-relation width of the $2S$ state is about one $Hz$.  A recent measurement gave \cite{Parthey11}
\be
f_{1S-2S}=2\,466\,061\,413\,187\,035(10) Hz
\ee
for the transition frequency with a precision corresponding to a fractional uncertainty of $4.2 \times 10^{-15}$.

We can easily use our results to evaluate the SME corrections to S state energies and splittings.  For $S$ states the momentum product $\hat p^i \hat p^j$ is effectively $\frac{1}{3} \delta_{i j}$ and the expectation values of \eqref{f_expec} and \eqref{f_expec_p} become
\bse \label{S_mes}
\bearray 
\langle m_f \vert \hat p^i \hat p^j \vert m_f \rangle_{+, \pm'} &=& \frac{1}{3} \delta_{i j} , \\
\langle m_f \vert \hat p^i \hat p^j \sigma_e^k \vert m_f \rangle_{+, \pm'} &=& \frac{1}{3} \delta_{i j} \xi_e^{\ell=0} \delta_{k 3} \, m_f =  \frac{1}{3} \delta_{i j} \delta_{k 3} \, m_f , \\
\langle m_f \vert \hat p^i \hat p^j \sigma_p^k \vert m_f \rangle_{+, \pm'} &=& \frac{1}{3} \delta_{i j} \xi_p^{\ell=0} \delta_{k 3} \, m_f =   \frac{1}{3} \delta_{i j} \delta_{k 3} \, m_f 
\eearray
\ese
where $\pm \rightarrow +1$, $j \rightarrow 1/2$, and $\pm' \rightarrow +1$ for states with $f=1$ (which states are the only ones with non-vanishing $m_f$).  It follows from \eqref{Coef_defs}, \eqref{fermion_result1}, \eqref{E_F_result1}, \eqref{E_AF}, and \eqref{S_mes}, or directly from \eqref{E2_result}, that the SME energy correction at $O(SME \cdot \alpha^2)$ is
\be
E^2_{\rm SME} \rightarrow \left \{ -\frac{5}{6} \left ( \tilde c^e_{0 0} + \epsilon \tilde c^p_{0 0} \right ) - \frac{2}{3} \tilde \kappa_0 + \frac{m_f}{3} \left ( \tilde d^e_3 + \epsilon \tilde d^p_3 - \tilde g^e_{D 3} - \epsilon \tilde g^p_{D 3} \right ) \right \} \left ( \frac{\alpha}{n} \right )^2 
\ee
for a state with a particular value of $n$.  The $2S-1S$ transition in hydrogen is dominated by two-photon emission that is subject to the selection rules: $\Delta f=0$, $\Delta m_f=0$. \cite{Cagnac73} It follows that there is no SME contribution to the $2S-1S$ energy splitting at $O(SME \cdot \alpha^0)$.  The leading correction to the splitting is at $O(SME \cdot \alpha^2)$: 
\be
\Delta E_{\rm SME}(2S-1S) = \left \{ \frac{5}{2} \left ( \tilde c^e_{0 0} + \epsilon \tilde c^p_{0 0} \right ) + 2 \tilde \kappa_0 - m_f \left ( \tilde d^e_3 + \epsilon \tilde d^p_3 - \tilde g^e_{D 3} - \epsilon \tilde g^p_{D 3} \right ) \right \} \frac{\alpha^2}{4} .
\ee
These results are consistent with $E^2_{SME} \rightarrow -\frac{m_e \alpha^2}{2 n^2} \left ( \frac{5}{3} c^e_{0 0} \right )$ of Altschul \cite{Altschul10}, who discarded coefficients other than $c^e_{\mu \nu}$ in \eqref{SME_coefs} as already well-bounded and did not consider the $k_F$ contribution, and with $\Delta \nu \rightarrow -\frac{\alpha^2 b^e_3}{8 \pi}$ for the difference between the $f=1, m_f=1$ and the $f=0,m_f=0$ frequency shifts of the $2S \rightarrow 1S$ transition, calculated as an example by Bluhm, Kosteleck\'y, and Russell \cite{Bluhm99}.

%%%%%%%%%%%%%%%%%%%%%%%%%%%%%%%%%%%%%%%%%%%%%%%%%%%

\section{Discussion}
\label{discussion}

Having values for the $O(SME \cdot \alpha^2)$ corrections to the hydrogen energy levels makes it possible to consider tests in hydrogen for additional SME parameters besides just $B^e_k$ and $B^p_k$.  In fact, $B^e_k$ and $B^p_k$ cancel entirely for the $2S-1S$ energy difference, leaving the $O(SME \cdot \alpha^2)$ contributions as the leading SME corrections.  In addition, it would be easy to find the corresponding corrections for anti-hydrogen, since the SME coefficients for antiparticles are related in a simple way to the coefficients for the corresponding particles.  Specifically, the coefficients for positrons and anti-protons are the same in magnitude as the coefficients for electrons and protons, but the $a^\mu$, $d^{\mu \nu}$, and $H^{\mu \nu}$ coefficients change sign. \cite{Kostelecky99a,Bluhm99}  Comparisons between hydrogen and anti-hydrogen have the potential to probe separately the various parts of $B_k$ and to more easily distinguish between the effects of $\tilde d_k$, which changes sign for anti-hydrogen, and $\tilde c_{0 0}$, $\tilde c_Q$, and $\tilde g_{D k}$, which do not.  The effects of the proton parameters are suppressed in hydrogen by a factor of $\epsilon = (m_e/m_p)^2 \approx 3 \times 10^{-7}$ relative to electron effects because the typical proton speed squared in hydrogen ($v_p^2 \sim (\vert p \vert/m_p)^2 \sim (\alpha m_e/m_p)^2 \sim 1.6 \times 10^{-11}$) is so much less than that of the electron ($v_e^2 \sim (\vert p \vert/m_e)^2 \sim \alpha^2 \sim 5 \times 10^{-5}$).  Much larger proton (and neutron) effects can be found in systems with a nucleus containing more than just a proton, as nucleon speeds in nuclei are much larger, typically $v^2 \sim 10^{-2}$. \cite{Kostelecky99b}

We expect that there are additional energy corrections of $O(SME \cdot \alpha^2)$ in hydrogen due to interactions such as an SME fermion line correction of the form $A + \vec B \cdot \vec \sigma$ combined with exchange of, say, a transverse photon.  Such un-calculated corrections depend on the same SME parameters as in \eqref{E0_result} but are higher order in $\alpha$.  We have restricted our attention to the leading order effects of any particular SME parameter.

In our work up until now we have assumed the existence of a small external magnetic field.  This field has the effect of breaking the degeneracy over $m_f$ so that the states $\vert m_f \rangle_{\pm, \pm'}$ of \eqref{f_states} are non-degenerate and non-degenerate perturbation theory can be used in our calculations.  We note that the direction of this external field determines which component of the relevant SME parameters can be probed by the hydrogen experiments.  The situation changes if this field is completely absent or shielded so well that its effect is less than that of the SME.  In that case, the degeneracy over $m_f$ would be broken by the $O(SME \cdot \alpha^0)$ perturbations, leading to a lowest order SME energy shift of
\be
E^0_{\rm SME} \rightarrow \langle m_f \vert \left ( A^e + A^p + B^e_k \sigma_e^k + B^p_k \sigma_p^k \right ) \vert m_f \rangle_{\pm, \pm'} = \left ( A^e + A^p \right ) + \left \vert \xi_e \vec B^e+\xi_p \vec B^p \right \vert m_f ,
\ee
where now the $z$ axis is determined by the direction of $\xi_e \vec B^e + \xi_p \vec B^p$.  The $O(SME \cdot \alpha^2)$ corrections take the same form as \eqref{E2_result} except that now the $z$ axis is determined by the SME vector $\xi_e \vec B^e + \xi_p \vec B^p$ instead of an external magnetic field.

We note that frequency measurements are always done relative to some standard.  For example, in the $2S \rightarrow 1S$ measurement of Parthey {\it et al.\/} \cite{Parthey11}, the measurement was done relative to a cesium fountain atomic clock.  A SME effect on hydrogen is in principle only detectable if it affects hydrogen and the reference clock differently.  In our case, the main SME effects for hydrogen are through $B^e_k$, $B^p_k$, the photon coefficients $\tilde \kappa_0$, $\tilde \kappa_Q$, and $(k_{AF})^k$, and the electron coefficients $c^e_{0 0}$, $\tilde c^e_Q$, $\tilde d^e_k$, and $\tilde g^e_{D k}$ (assuming that the tiny $\epsilon$ factor causes the proton contribution at $O(SME \cdot \alpha^2)$ to be negligible), while nuclear effects are important in cesium.  ``Clock comparison'' experiments have been discussed extensively by Bluhm {\it et al.\/} in the context of the SME. \cite{Bluhm02,Bluhm03}

%%%%%%%%%%%%%%%%%%%%%%%%%%%%%%%%%%%%%%%%%%%%%%%%%%%

\begin{acknowledgments}
We acknowledge the support  of Franklin \& Marshall College through the Hackman Scholars program.
We are grateful to Quentin Bailey, Richard Fell, Michael Hohensee, Ralf Lehnert, Neil Russell, and Calvin Stubbins for discussions, and to an anonymous referee for pointing out the issues related to coordinate transformations.
\end{acknowledgments}

%%%%%%%%%%%%%%%%%%%%%%%%%%%%%%%%%%%%%%%%%%%%%%%%%%%

\appendix

\section{Effect of a coordinate transformation}

Many SME coefficients or combinations of coefficients are unphysical and cannot be observed.  Unobservable fermion coefficients, such as $a_\mu+m e_\mu$, $c_{\mu \nu}-c_{\nu \mu}$, and $f_\mu$, along with the corresponding field redefinitions used in the demonstrations of unobservability, have been discussed by many authors. \cite{Colladay98,Colladay02,Altschul06,Fittante12} Invariance of the physics under particular coordinate transformations can be used to obtain relationships among certain photon and fermion coefficients, specifically between a subset of the $(k_F)_{\mu \nu \rho \sigma}$ photon coefficients and the symmetric and traceless $c_{\mu \nu}$ fermion coefficients. \cite{Colladay98,Kostelecky02b,Muller03,Bailey04,Hohensee09,Kostelecky11b} Suppose we consider a theory containing only electrons and photons, and only the $c_{\mu \nu}$ and $(k_F)_{\mu \nu \rho \sigma}$ SME coefficients: the Lagrangian reduces to
\be
{\cal L} = \frac{1}{2} i \bar \psi \gamma_\mu \left ( \eta^{\mu \nu}+c^{\mu \nu} \right ) \! \! \stackrel{\;\leftrightarrow}{D}_{\nu} \! \psi - \bar \psi m \psi - \frac{1}{4} F_{\mu \nu} \left ( \eta^{\mu \rho} \eta^{\nu \sigma} + (k_F)^{\mu \nu \rho \sigma} \right ) F_{\rho \sigma} .
\ee
The nineteen independent coefficients of $(k_F)^{\mu \nu \rho \sigma}$ have the symmetries of the Riemann tensor and are double traceless by convention: \cite{Colladay98,Kostelecky02b}
\bse
\bearray
(k_F)_{\mu \nu \rho \sigma} = (k_F)_{\rho \sigma \mu \nu} &=& - (k_F)_{\nu \mu \rho \sigma} , \\
(k_F)_{\mu \nu \rho \sigma} + (k_F)_{\mu \rho \sigma \nu} + (k_F)_{\mu \sigma \nu \rho} &=& 0 , \\
(k_F)^{\mu \nu}_{\phantom{\mu \nu} \mu \nu} &=& 0 .
\eearray
\ese
The $(k_F)^{\mu \nu \rho \sigma}$ tensor can be decomposed into a ten-component completely traceless part $C^{\mu \nu \rho \sigma}$ and a nine-component part defined in terms of a traceless tensor $\tilde k^{\mu \nu} \equiv (k_F)^{\mu \alpha \nu}_{\phantom{\mu \alpha \nu} \alpha}$ according to: \cite{Bailey10,Kostelecky04,Kostelecky09}
\be \label{decomposition.of.kF}
(k_F)^{\mu \nu \rho \sigma} = C^{\mu \nu \rho \sigma} + \frac{1}{2} \left ( \eta^{\mu \rho} \tilde k^{\nu \sigma} - \eta^{\nu \sigma} \tilde k^{\nu \rho} + \eta^{\nu \sigma} \tilde k^{\mu \rho} - \eta^{\nu \rho} \tilde k^{\mu \sigma} \right ) .
\ee
We restrict our model further to have $C^{\mu \nu \rho \sigma}=0$.  The coordinate transformation
\be \label{transform}
x^\mu \rightarrow x'^\mu = x^\mu - \frac{1}{2} \tilde k^\mu_{\phantom{\mu} \nu} x^\nu = \left ( \delta^\mu_{\phantom{\mu} \nu} - \frac{1}{2} \tilde k^\mu_{\phantom{\mu} \nu} \right ) x^\nu = \frac{\partial x'^\mu}{\partial x^\nu} x^\nu
\ee
has unit Jacobian due to the tracelessness of $\tilde k^\mu_{\phantom{\mu} \nu}$ and allows us to write ${\cal L}$ as
\be
{\cal L} = \frac{1}{2} i \bar \psi' \gamma_\mu \left ( \eta^{\mu \nu}+c^{\mu \nu} - \frac{1}{2} \tilde k^{\mu \nu} \right ) \! \! \stackrel{\;\leftrightarrow}{D'}_{\nu} \! \psi' - \bar \psi' m \psi' - \frac{1}{4} F'_{\mu \nu} \eta^{\mu \rho} \eta^{\nu \sigma} F'_{\rho \sigma}
\ee
where $\psi'=\psi(x')$, $D'_\nu = \partial'_\nu + i q A'_\nu(x')$, etc.  So a theory of electrons with the usual SME $c_{\mu \nu}$ coefficient and photons with the SME $(k_F)_{\mu \nu \rho \sigma}$ coefficient given by the $\tilde k^{\mu \nu}$ part of \eqref{decomposition.of.kF} is equivalent to a theory of electrons with $c_{\mu \nu} \rightarrow c_{\mu \nu}-(1/2) \tilde k_{\mu \nu}$ and a conventional quadratic photon Lagrangian.
This equivalence must be evident in the electron-photon part of our result for hydrogen energy levels.  However, we do not expect the 
energy in the unprimed coordinate system to equal the energy in the primed coordinate system--instead, it should be related according to transformation \eqref{transform}.  For a state with energy $E$ and time dependence $\exp \left ( -i E t \right )$ in the unprimed system, the time dependence in the primed system takes the form $\exp \left ( -i E (1+\tilde k_{0 0}/2) t' \right )$.  The corresponding primed system energy will be
\be \label{energy_transform}
E' = \left (1 + \frac{1}{2} \tilde k_{0 0} \right ) E.
\ee
The relevant part of our hydrogen energy result \eqref{E2_result} is
\bearray \label{unprime.energy}
E &=& m \left \{ 1 - c_{0 0} + \left ( \frac{\alpha}{n} \right )^2 \left [ -\frac{1}{2} - \frac{5}{6} c_{0 0} - \frac{2}{3} (k_F)_{a 0 a 0} + X \delta^q_{i j} c_{i j}  - X \delta^q_{i j} (k_F)_{i 0 j 0} \right ] \right \} \cr
&=& m \left \{ 1 - c_{0 0} + \left ( \frac{\alpha}{n} \right )^2 \left [ -\frac{1}{2} - \frac{5}{6} c_{0 0} + \frac{2}{3} \tilde k_{0 0} + X \delta^q_{i j} \left ( c_{i j} - \frac{1}{2} \tilde k_{i j} \right ) \right ] \right \}
\eearray
where $X$ is a relatively complicated function of the angular quantum numbers and \eqref{decomposition.of.kF} with $C^{\mu \nu \rho \sigma}=0$ leads to
\be
(k_F)_{i 0 j 0} = \frac{1}{2} \left ( \tilde k_{i j} - \delta_{i j} \tilde k_{0 0} \right )
\ee
so that $\delta_{i j} (k_F)_{i 0 j 0} = (k_F)_{a 0 a 0} = - \tilde k_{0 0}$ and $\delta^q_{i j} (k_F)_{i 0 j 0} = \frac{1}{2} \delta^q_{i j} \tilde k_{i j}$.  We transform to the primed system by first making the replacement $\tilde k_{\mu \nu} \rightarrow 0$ followed by the replacement $c_{\mu \nu} \rightarrow c_{\mu \nu}-(1/2) \tilde k_{\mu \nu}$ in \eqref{unprime.energy}, and obtain
\be
E' = m \left \{ 1 - \left ( c_{0 0} - \frac{1}{2} \tilde k_{0 0} \right ) + \left ( \frac{\alpha}{n} \right )^2 \left [ -\frac{1}{2} - \frac{5}{6} \left ( c_{0 0} - \frac{1}{2} \tilde k_{0 0} \right ) + X \delta^q_{i j} \left ( c_{i j} - \frac{1}{2} \tilde k_{i j} \right ) \right ] \right \}.
\ee
To first order in SME coefficients this has the required form \eqref{energy_transform}.

\break

%%%%%%%%%%%%%%%%%%%%%%%%%%%%%%%%%%%%%%%%%%%%%%%%%%%

%%%%%%%%%%%%%%%%%%%%%%%%%%%%%%%%%%%%%%%%%%%%%%%%%%%


\section*{References}
\begin{thebibliography}{10}

\bibitem{Mohr08} P. J. Mohr, B. N. Taylor, and D. B. Newell, Rev. Mod. Phys. {\bf 80}, 633 (2008).
\bibitem{Robilliard11} Many recent advances in hydrogen physics have been reported in the Precision Physics of Simple Atomic Systems series of meetings.  The most recent was held in 2012.  The proceedings of the 2010 meeting were published in the Canadian Journal of Physics: see C. Robilliard and S. G. Karshenboim, Can. J. Phys. {\bf 89}, v (2011).
\bibitem{Eides01} M. I. Eides, H. Grotch, and V. A. Shelyuto, Phys. Reports {\bf 342}, 63 (2001).
\bibitem{Karshenboim05} S. G. Karshenboim, Phys. Reports {\bf 422}, 1 (2005).
\bibitem{Colladay97} D. Colladay and V. A. Kosteleck\'y, Phys. Rev. D {\bf 55}, 6760 (1997).
\bibitem{Colladay98} D. Colladay and V. A. Kosteleck\'y, Phys. Rev. D {\bf 58}, 116002 (1998).
\bibitem{Kostelecky95} V. A. Kosteleck\'y and R. Potting, Phys. Rev. D {\bf 51}, 3923 (1995).
\bibitem{Kostelecky11} V. A. Kosteleck\'y, and N. Russell, Rev. Mod. Phys. {\bf 83}, 11 (2011).
\bibitem{CPT} V. A. Kosteleck\'y, ed., {\it CPT and Lorentz Symmetry I, II, III, IV, V\/}, (World Scientific, Singapore, 1999, 2002, 2005, 2008, 2011).
\bibitem{Bluhm06} R. Bluhm, Lect. Notes Phys. {\bf 702}, 191 (2006) [arXiv:hep-ph/0506054].
\bibitem{Lehnert06} R. Lehnert, arXiv:hep-ph/0611177v2 (2006).
\bibitem{Bluhm99} R. Bluhm, V. A. Kosteleck\'y, and N. Russell, Phys. Rev. Lett. {\bf 82}, 2254 (1999).
\bibitem{Shore05} G. M. Shore, Nucl. Phys. B {\bf 717}, 86 (2005).
\bibitem{Ferreira06} M. M. Ferreira, Jr. and F. M. O. Moucherek, Int. J. Mod. Phys. A {\bf 21}, 6211 (2006).
\bibitem{Kharlanov07} O. G. Kharlanov and V. C. Zhukovsky, J. Math. Phys. {\bf 48}, 092302 (2007).
\bibitem{Altschul10} B. Altschul, Phys. Rev. D {\bf 81}, 041701(R) (2010).
\bibitem{Belich06} H. Belich, T. Costa-Soares, M. M. Ferreira, Jr., J. A. Helay\"el-Neto, and F. M. O. Moucherek, Phys. Rev. D {\bf 74}, 065009 (2006).
\bibitem{Kostelecky99a} V. A. Kosteleck\'y and C. D. Lane, J. Math. Phys. {\bf 40}, 6245 (1999).
\bibitem{Kostelecky99b} V. A. Kosteleck\'y and C. D. Lane, Phys. Rev. D {\bf 60}, 116010 (1999).
\bibitem{Lehnert04} R. Lehnert, J. Math. Phys. {\bf 45}, 3399 (2004).
\bibitem{Parthey11} C. G. Parthey {\it et al.\/}, Phys. Rev. Lett. {\bf 107}, 203001 (2011).
\bibitem{Kostelecky02a} V. A. Kosteleck\'y, C. D. Lane, and A. G. M. Pickering, Phys. Rev. D {\bf 65}, 056006 (2002).
\bibitem{Itzykson80} C. Itzykson and J.-B. Zuber, {\it Quantum Field Theory} (McGraw-Hill, New York, 1980).
\bibitem{Foldy50} L. L. Foldy and S. A. Wouthuysen, Phys. Rev. {\bf 78}, 29 (1950).
\bibitem{Colladay02} D. Colladay and P. McDonald, J. Math. Phys. {\bf 43}, 3554 (2002).
\bibitem{Altschul06} B. Altschul, J. Phys. A: Math. Gen. {\bf 39}, 13757 (2006).
\bibitem{Fittante12} A. Fittante and N. Russell, arXiv:1210.2003v1.
\bibitem{Edmonds57} See, for example, A. R. Edmonds, {\it Angular Momentum in Quantum Mechanics\/}, (Princeton U. Press, Princeton, 1957).\bibitem{Bluhm02} R. Bluhm, V. A. Kosteleck\'y, C. D. Lane, and N. Russell, Phys. Rev. Lett. {\bf 88}, 090801 (2002).
\bibitem{Bluhm03} R. Bluhm, V. A. Kosteleck\'y, C. D. Lane, and N. Russell, Phys. Rev. D {\bf 68}, 125008 (2003).
\bibitem{Podolsky29} B. Podolsky and L. Pauling, Phys. Rev. {\bf 34}, 109 (1929).
\bibitem{Fock36} V. Fock, Z. Phys. {\bf 98}, 145 (1936).
\bibitem{Kinoshita96} T. Kinoshita and M. Nio, Phys. Rev. D {\bf 53}, 4909 (1996).
\bibitem{Labelle98} P. Labelle, Phys. Rev. D {\bf 58}, 093013 (1998).
\bibitem{Berestetskii} V. B. Berestetskii, E. M. Lifshitz, and L. P. Pitaevskii, {\it Quantum Electrodynamics\/}, Vol. 4 of the Landau and Lifshitz Course of Theoretical Physics, 2$^{nd}$ ed. (Pergamon Press, Oxford, 1982), Sec.~83.
\bibitem{Kostelecky02b} V. A. Kosteleck\'y and M. Mewes, Phys. Rev. D {\bf 66}, 056005 (2002).
\bibitem{Bailey04} Q. G. Bailey and V. A. Kosteleck\'y, Phys. Rev. D {\bf 70}, 076006 (2004).
\bibitem{Bailey10} Q. G. Bailey, Phys. Rev. D {\bf 82}, 065012 (2010).
\bibitem{Breit40} G. Breit and E. Teller, Astrophys. J. {\bf 91}, 215 (1940).
\bibitem{Klarsfeld69} S. Klarsfeld, Phys. Lett. A {\bf 30}, 382 (1969).
\bibitem{Cagnac73} B. Cagnac, G. Grynberg, and F. Biraben, J. Phys. {\bf 34}, 845 (1973).
\bibitem{Muller03} H. M\"uller, S. Herrmann, A. Saenz, A. Peters, and C. L\"ammerzahl, Phys. Rev. D {\bf 68}, 116006 (2003).
\bibitem{Hohensee09} M. A. Hohensee, R. Lehnert, D. F. Phillips, and R. L. Walsworth, Phys. Rev. D {\bf 80}, 036010 (2009).
\bibitem{Kostelecky11b} V. A. Kosteleck\'y and J. D. Tasson, Phys. Rev. D {\bf  83}, 016013 (2011).
\bibitem{Kostelecky04} V. A. Kosteleck\'y, Phys. Rev. D {\bf 69}, 105009 (2004).
\bibitem{Kostelecky09} V. A. Kosteleck\'y and M. Mewes, Phys. Rev. D {\bf 80}, 015020 (2009).
\end{thebibliography}
\end{document}